# Simple methods for estimating confidence levels, or tentative probabilities, for hypotheses instead of *p* values

**Michael Wood**

**Abstract**

In many fields of research, null hypothesis significance tests and *p* values are the accepted way of assessing the degree of certainty with which research results can be extrapolated beyond the sample studied. However, there are very serious concerns about the suitability of *p* values for this purpose. An alternative approach is to cite confidence intervals for a statistic of interest, but this does not directly tell readers how certain a hypothesis is. Here, I suggest how the framework used for confidence intervals could easily be extended to derive confidence levels, or "tentative probabilities," for hypotheses. I also outline four quick methods for estimating these. This allows researchers to state their confidence in a hypothesis as a direct probability, instead of circuitously by *p* values referring to a hypothetical null hypothesis—which is usually not even stated explicitly. The inevitable difficulties of statistical inference mean that these probabilities can only be tentative, but probabilities are the natural way to express uncertainties, so, arguably, researchers using statistical methods have an obligation to estimate how probable their hypotheses are by the best available method. Otherwise, misinterpretations will fill the void.

## Introduction

There have been extensive criticisms of null hypothesis significance tests and *p* values in the literature for more than 50 years (e.g. Morrison and Henkel, 1970; Nickerson, 2000; Nuzzo, 2014; Wasserstein and Lazar, 2016): they are widely misinterpreted, and do not, as is often assumed, give the probability of a hypothesis being true, nor a measure of the size or importance of the effect.

To take an example, Redelmeier and Singh (2001) concluded that "life expectancy was 3.9 years longer for Academy Award [Oscar] winners that for other, less recognized performers (79.7 vs. 75.8 years, $p=0.003$)" (p. 955). The *p* value gives an indication of how susceptible the result may be to sampling error, but only for readers who understand the role of the unstated null hypothesis.

The statistically naive reader may jump to a number of misleading conclusions from the *p* value of 0.003. This might be assumed to be the probability of the effect being due to chance factors, so the probability of winning an Oscar really prolonging the actor's life might be taken to be $1-0.003$ or 99.7%, which is an impressive degree of certainty. The use of the word "significant" may lead to the assumption that the conclusion is big or important. In this case readers are told the difference between winners and non-winners is 3.9 years, but in many published papers the size of the effect is not given. And just the fact that the *p* value is well below the conventional threshold of 5% may persuade readers that the result is beyond reasonable doubt.

However, all of these conclusions are wrong. Using the approach to be explained below the "confidence level" for Oscar winners living longer is 99.85%, and the confidence level for their living more than a year longer is 98.6%. This avoids the use of the word "significant" with its potentially misleading connotations, and gives the opportunity to give

University of Portsmouth, Portsmouth, UK

**Corresponding author:**
Michael Wood, University of Portsmouth, Richmond Building, Portland Street, Portsmouth PO1 2UP, UK.
Emails: michaelwoodslg@gmail.com; michael.wood@port.ac.uk

confidence levels for the extra longevity of Oscar winners being more than a specified number of years. *p* values, correctly interpreted, simply do not tell us how likely the hypothesis of increased longevity for Oscar winners is. Which is, of course, what we really want to know.

Furthermore, the fact that the result is well below the conventional cutoff value of 5%, and so "very significant," may encourage readers to see the result as more conclusive than is realistic. The *p* value in this example only analyzes sampling errors on the assumption that the sample of past Oscar winners is a random sample of potential Oscar winners in the past and future—a tricky assumption to make sense of, let alone test. Other potential sources of uncertainty, like the way the data was collected, changes which may occur over time, and mistakes in the analysis of data are ignored. In the present case, this last possibility was an issue: the conclusions of this study were challenged by Sylvestre et al. (2006) in a later article in the same journal. The challenge had nothing to do with the *p* values, so the details are not relevant here, but it does illustrate that the assumption that a low *p* value indicates a credible result needs to be treated with caution. (There is a clue about the problem in the way I have slid from the idea of Oscar winners living longer to the idea of winning an Oscar *causing* actors to live longer as opposed to living longer giving actors more time to win Oscars.)

The American Statistical Association's "statement on *p*-values" (Wasserstein and Lazar, 2016) proposes six principles, of which five are negative in the sense that they include the word "not" or "incompatible," and the remaining one is a platitude ("Proper inference requires full reporting and transparency"). For example, the second is "p-values do not measure the probability that the studied hypothesis is true, or the probability that the data were produced by random chance alone." They go on to mention, very briefly, some other approaches, and conclude with another negative sentence: "No single index should substitute for scientific reasoning." This all puts the researcher in a difficult position. They are told what not to assume, but little about what they can believe.

My aim here is not to add to this critical, but often ineffective, literature on *p* values (interested readers will find many references to this literature in the articles cited above), but to suggest a simple alternative to *p* values which has a clear interpretation and would not need to be hedged in by so many caveats to reduce the likelihood of misinterpretation. Furthermore, it does answer the important question: how likely is it that this hypothesis is true? On the other hand it is a single index, and I would concur with the ASA's statement that "no single index should substitute for scientific reasoning." My proposal is simply to use confidence distributions (Xie and Singh, 2013), which are the basis of confidence intervals, to estimate confidence levels for whatever hypotheses are of interest. Instead of giving a *p* value (e.g. $p=4\%$) to indicate our degree of confidence in a statistical conclusion, we could give a confidence level for the conclusion (e.g. $CL=98\%$). This is far more straightforward and informative than the *p* value, but it is important to be aware that these confidence levels can only be tentative. This tentativeness applies equally to confidence intervals, although this is rarely acknowledged. Sometimes, of course, *p* values are a sensible approach (see Wood, 2016 and section "How should we describe the strength of the evidence based on a sample: confidence levels or *p* value" below), but often, perhaps usually, they are not.

There are a variety of methods of deriving confidence distributions: according to a recent review the "concept of a confidence distribution subsumes and unifies a wide range of examples, from regular parametric (fiducial distribution) examples to bootstrap distributions, significance (*p* value) functions, normalized likelihood functions, and, in some cases, Bayesian priors and posteriors" (Xie and Singh, 2013: 3). To review all these would take me too far from my main concern, but I will explain how estimates of confidence levels can be obtained from the *p* values and confidence intervals produced by computer packages, from bootstrapping, and the relationship with Bayes' theorem.

I will also suggest that the distinction between confidence and probability is an unnecessary complication, and that confidence levels should be viewed as "tentative probabilities."

Given a confidence distribution, the mathematics of my proposal is trivial and best explained by means of an example, although the scope of the approach is far wider than this example. My aim here is to suggest how a well-established statistical framework could easily be extended to make it far more useful for research by providing a straightforward answer to the obvious question: "How likely is it that this hypothesis is true?" (As we have seen, *p* values do not do this.)

In the next section, I introduce the example and how it might be analyzed conventionally, and then in the section "Confidence levels for hypotheses," I use this to give a brief introduction to the standard approach to confidence intervals and distributions, and explain how this can be used to derive confidence levels for more general hypotheses. In the section "Tentative probabilities or confidence levels?," I discuss the link with Bayes' theorem, and whether it is worth preserving the distinction between confidence and probability. The section "Methods for quick estimates of tentative probabilities, or confidence levels, for hypotheses" outlines several methods for estimating confidence levels: these include bootstrapping, which is a newer approach to the idea of confidence. The section "How should we describe the strength of the evidence based on a sample: confidence levels or *p* values" discusses some more examples—including one in which *p* values are a sensible approach.

## An example

I will use the fictional "data" in Table 1 to illustrate the approach. These are scores from random samples from two populations, A and B. (The scores might be from a psychological test, or measures of the effectiveness of two treatments; the samples might be from two countries or different treatment groups in a randomized trial.) One of the researcher's hypotheses was that the mean score in one of the populations (A) would be higher than in the other (B). The difference between the means is small (0.3), and with the small sample of 10 in each population, the $p$ value is high (0.673)[1] and the 95% confidence interval for the difference (−1.2 to +1.8) encompasses both positive and negative values, indicating that we cannot be sure which population has the higher mean score. The $p$ values and confidence intervals are based on the standard methods with the $t$ distribution, using the independent samples $t$ test in SPSS. The data, and the formula for the $t$ test, are in the online supplement material spreadsheet popsab.xlsx.

**Table 1.** Data and conventional analysis for small and large samples.

| | | |
|---|---|---|
| ***Data and analysis for small samples (n=10 from each population)*** | | |
| | Sample from Population A | Sample from Population B |
| | 8 | 8 |
| | 5 | 7 |
| | 6 | 5 |
| | 6 | 6 |
| | 6 | 3 |
| | 5 | 8 |
| | 8 | 6 |
| | 7 | 6 |
| | 7 | 3 |
| | 5 | 8 |
| Means: | 6.3 | 6 |
| $p$ value (two tailed) for the difference of means: | | 0.673 (67.3%) |
| 95% Confidence interval: | | −1.2 to +1.8 |
| ***Large samples comprising 40 copies of small samples (n=400 from each pop)*** | | |
| Obviously the means are the same as for the small samples | | |
| $p$ value (two tailed) for the difference of means: | | 0.004 (0.4%) |
| 95% Confidence interval: | | +0.1 to +0.5 |

Ten is obviously a very small sample, but it is helpful to show the contrast with a larger sample of 400 from each population. This is constructed by taking 40 copies of the small sample so that the distribution is identical to that of the small sample. This large sample is obviously not a realistic sample (there are, for example, 40 fives and 80 threes but no fours in the sample from Population A which is most unlikely in practice), but it does serve the purpose of illustrating the contrast between samples of different sizes from the same population. I have also included a more realistic large sample in the online supplement material spreadsheet, popsab.xlsx. This has the same sample sizes and the same distribution in the sense of the same means and standard deviations for each population, as the 40 copies sample, so obviously the $t$ test and confidence interval results are also identical, because these just depend on the mean and standard deviation in each sample.

With the larger sample (see Table 1), the difference is significant at the 1% level, and the confidence interval is entirely in the positive range indicating that the mean score in Population A is likely to be more than that in B. However, the apparent strength of these conclusions makes it easy to forget that the estimated difference between the population means is only 0.3.

## Confidence levels for hypotheses

The confidence interval for the difference of the means based on the small samples can be derived from the confidence distribution in Figure 1 (bold curve). It should be roughly obvious that 95% of the "confidence" lies between −1.2 and +1.8 (the limits of the confidence interval in Table 1), with 2.5% in each of the tails since the distribution is symmetrical. (The vertical lines, and the phrase "tentative probability," are explained below.) Similarly the 50% confidence interval would be two lines which enclose 50% of the area between them.

It is useful to see the link between confidence intervals and $p$ values because this shows how confidence intervals can be derived. Smithson (2003: 78) defines a "plausible" value for a population statistic as one that lies within the confidence interval, and an implausible one as one that lies outside it. So, based on the small sample and the 95% interval (−1.2 to +1.8), plausible values for the difference of the population means are those that lie between −1.2 and +1.8 and implausible values are those that lie outside the interval. In terms of significance tests and $p$ values, if the null hypothesis value of the parameter is implausible this will mean that $p <= 5\%$ (100%–95%), and if the null hypothesis value is on the boundary of the confidence interval (−1.2 or +1.8) then $p = 5\%$. So, to find the ends of a 95% confidence interval, we need to find the null hypothesis values which will give a (two tail) $p$ value of 5%: in this case these values are −1.2 and +1.8. These are obviously the 2.5 and 97.5 percentiles of the confidence distribution. The rest of the confidence distribution can be filled in similarly by considering other confidence intervals and $p$ values. This is the basis of the graphs in Figure 1. (For a fuller explanation see Smithson (2003), or any other standard text.)

Conventionally, the word "confidence" is only used in relation to intervals, usually 95% ones. In practical terms this is odd because the 95% is based on arbitrary statistical convention, rather than the requirements of the problem. My suggestion here is that confidence distributions could be used to derive confidence levels for more general hypotheses.

Figure 1 can be used to derive a confidence level for the mean score in Population A being more than in B—that is, the difference of the means being positive. This is the proportion of the area to the right of the line representing a difference of zero. To derive this from a confidence interval, we need one of the limits for the interval to be zero. This can be estimated by trial and error with a package such as SPSS: the required confidence interval for the small samples is the 32.7% interval (0–0.6), which leaves 33.65% in each of the tails, and the "confidence" to the right of the solid vertical line is 66.35% (32.7% + 33.65%)—which is our confidence that the mean score of Population A is greater than that of Population B. Alternatively, the same result could be obtained directly from the formulae used to draw Figure 1 (see online supplement material spreadsheet popsab.xlsx).

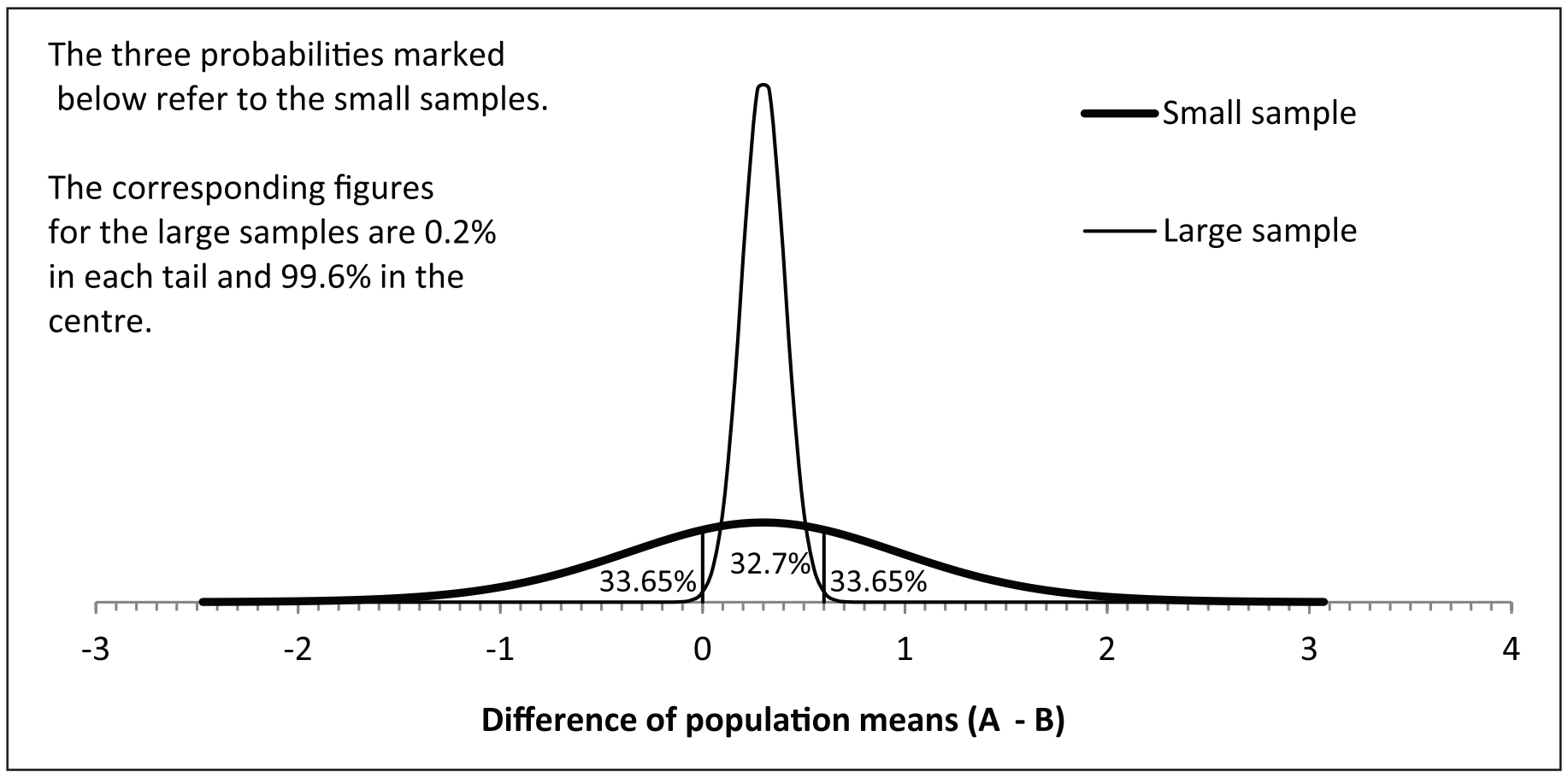

**Figure 1.** Confidence/tentative probability distributions for the difference between the mean scores in two populations based on small and large samples.

This principle can easily be extended. For example, we might decide that small differences between the populations are not of interest so we might have three hypotheses: the mean scores in the two populations are within one unit of each other, A is substantially more than B (more than one unit more), and vice versa. Table 2 brings all these results together for both the small and big samples in Table 1.

Table 2 shows the obvious contrast between the small and large samples. For the large sample the evidence is almost conclusive (100.0%) that A has *approximately* the same mean score as B, whereas for the small sample this hypothesis only has a 79% confidence. All the probabilities for the large sample are close to 0% or 100%—which indicates the greater level of certainty obtained by taking larger samples. Figure 1 also shows this contrast: the confidence distribution for the large sample is much less spread out than the distribution for the small sample.

The large sample result illustrates one of the problems with $p$ values. The low $p$ value (0.004) suggests that the null hypothesis of *exactly* the same mean scores in both populations should be rejected. Population A does have a higher mean that Population B. However, the confidence level of 100.0% for the mean score in the two populations being *approximately* equal (within one unit) indicates that, although differences do exist between the two populations, they may be too small to matter in practice. Over-reliance on $p$ values may obscure this conclusion.

This approach would obviously work for any hypothesis based on a numerical statistic for which a confidence distribution like Figure 1 can be derived. Other standard examples include regression and correlation coefficients, and the difference between two proportions. Figure 1 is symmetrical, so the confidence level in each tail is obviously the same: where this is not the case the method would need to be adapted in an obvious way. The method can even be extended to hypotheses that are not defined by a single numerical statistic—see the section "Method 4: Bootstrapping" below, for an example.

**Table 2.** Confidence levels/tentative probabilities for hypotheses about how the mean score in Population A compares with Population B.

| Hypothesis about mean scores | Difference (A − B) | Confidence level or tentative probability (%) | Size of sample from each population |
|---|---|---|---|
| A > B | >0 | 66 | 10 |
| A < B | <0 | 34 | 10 |
| A >> B | >1 | 17 | 10 |
| A ≈ B | Between −1 and 1 | 79 | 10 |
| A << B | <−1 | 4 | 10 |
| A > B | >0 | 99.8 | 400 |
| A < B | <0 | 0.2 | 400 |
| A >> B | >1 | 0.0 | 400 |
| A ≈ B | Between −1 and 1 | 100.0 | 400 |
| A << B | <−1 | 0.0 | 400 |

The confidence levels/tentative probabilities are based on the distributions in Figure 1.
≈ means approximately equal which is defined as being within one unit.
>> means substantially more which is defined as more than a unit.

## Tentative probabilities or confidence levels?

According to the dominant, frequentist, version of statistical theory, confidence levels are *not* probabilities because there is just one true value of the statistic so probabilities are irrelevant. Probabilities apply to uncertain *events*, like whether a coin lands on heads or tails, *not* to beliefs or hypotheses (e.g. those in the left-hand column of Table 2), which are either true or false.

However, it is possible to interpret confidence levels within the Bayesian statistical paradigm. This uses Bayes' theorem and a prior probability distribution—reflecting our prior knowledge of the situation—to derive a posterior probability distribution on which "credible intervals" are based. If we make the neutral assumption of a flat prior distribution (all the values on the horizontal axis of Figure 1 are equally likely), then confidence intervals are often identical to Bayesian credible intervals. This is not true in general, but it is true for many distributions (including those in Figure 1) provided that the prior distribution is flat (Bolstad, 2007; Xie and Singh, 2013). Xie and Singh (2013) say this "appears to be a mathematical coincidence," which ignores the fact that such coincidences can be analyzed mathematically to uncover the circumstances in which they occur (see the Appendix of Wood, 2014).

This means that confidence distributions can be used "in the style of a Bayesian posterior" probability distribution (Xie and Singh, 2013), which means that confidence levels for hypotheses can be interpreted as Bayesian posterior probabilities. Bayesian posterior probability distributions *are* probability distributions: the main difference between the (standard) frequentist and Bayesian schools of probability theory being that that the former defines probability in terms of frequencies of events, whereas the latter has a more flexible definition involving some kind of degree of certainty. However, in practice, they both obey the same rules, and in everyday discourse the idea of probability is used in the broader, Bayesian sense. For example, if someone gives a probability for life existing on one of Jupiter's moons, this probability can only be interpreted as a degree of belief. And on the website of the medical journal BMJ (2017), confidence intervals are explained in these terms: "this is called the 95% confidence interval, and we can say that there is only a 5% chance that the range 86.96 to 89.04 mmHg excludes the mean of the population," which is a probabilistic explanation if we assume that "chance" means probability. For all these reasons, there seems little reason to distinguish between confidence and probability: using the term probability in both contexts would avoid the confusion of the extra term "confidence."

Despite this I will continue to use the term confidence in this article simply because of its familiarity: people know what confidence intervals are and using the term "confidence" taps into this knowledge. But I would argue that confidence levels should be regarded as *tentative* probabilities.

The qualifying adjective "tentative" stems from the inevitable difficulties of calculating probabilities for hypotheses. Imagine, with the data in Table 1, that we subsequently found that the data was a hoax and all samples came from the same population. This would mean that the mean population scores were identical and the observed difference was just sampling error for both big and small samples. This illustrates the principle that prior beliefs *must* have an impact on sensible conclusions: if we have a hypothesis which we are sure is false, no evidence will suffice to overturn it. This is where Bayesian statistics is helpful. A suitable prior distribution incorporating our prior beliefs will ensure that we get a sensible answer. On the other hand, if we have no definite prior information, a Bayesian interpretation of confidence levels has the advantage of yielding a probability, and of clarifying the main condition for the validity of the probability—namely that the prior distribution should be flat.

There is a plethora of other concepts in this area—fiducial probabilities, Bayes' factors, and so on. However, none of them are widely used, probably because they don't produce useful and intuitive measures of the certainty of a hypothesis. My contention here is that Bayes' theorem gives an answer in principle, but in practice we have to make simplifying assumptions about prior probabilities, and that extending the confidence interval framework is a good compromise.

## Methods for quick estimates of tentative probabilities, or confidence levels, for hypotheses

The general method is simply to use the confidence distribution used for confidence intervals to make the appropriate estimates, or to use Bayes' theorem as discussed in the section "Tentative probabilities or confidence levels?" above. However, this may not always be practicable. Even if we only have the confidence intervals or *p* values cited in a conventional analysis, it may still be possible to estimate confidence levels for the hypotheses of interest (Methods 2 and 3 below). And if we have the data, but do not know of a mathematical model for the confidence distribution, it may be possible to use the confidence interval routine built into statistical packages to reverse engineer the required confidence levels (Method 1), or to use bootstrapping to simulate a confidence distribution (Method 4). Bootstrapping is a very flexible simulation method: it can, for example, be used to estimate a confidence level for the relationship between two variables being an inverted U shape (see "Method 4: Bootstrapping" below and Wood, 2013: 7).

### *Method 1: using the confidence interval routine in statistical packages*

If the confidence intervals have equal confidence levels in each tail (this is usually the case), trial and error trying different confidence levels for the intervals can be used to estimate confidence levels for hypotheses as explained in the section "Confidence levels for hypotheses" above and illustrated by Figure 1.

### *Method 2: using* p *values*

Table 1 shows that the *p* value based on the big sample is 0.4%, and Table 2 shows that the confidence level for the mean of Population A being greater than B is 99.8%. This confidence level can be worked out from the *p* value using the relationship between *p* values and confidence levels explained in the section "Confidence levels for hypotheses" above. *p* values are conventionally based on a null hypothesis value of 0 (no difference between the two means), so one end of the 99.6% (100% − 0.4%) confidence interval must be 0. This means that the confidence level for the difference being below 0 is 0.2% (0.4%/2) and the confidence level for the difference being positive is 99.8%.

When the confidence distribution is a symmetrical curve like Figure 1, this leads to the following conclusions:

> Suppose we have a two tailed *p* value for a statistic based on a null hypothesis value of zero. Suppose further that the sample value of the statistic is positive but negative values would be possible. The difference of means in Table 1 is an example; other examples include correlation and regression coefficients, but not chi squared which is always positive. Then
>
> $$\textit{Confidence level that population value of statistic is positive} = 1 - p/2$$
>
> $$\textit{Confidence level that population value of statistic is negative} = p/2$$
>
> If the sample value is negative these would be reversed.
>
> These formulae can easily be adapted if the information we have about *p* is an inequality. For example if $p < 0.1\%$ then the first equation above becomes
>
> $$\textit{Confidence level that population value of statistic is positive} > 99.95\%$$

This method only works if the cutoff value for the hypotheses is zero. Method 3 is more general.

### *Method 3: using the normal or* t *distribution to extrapolate from* p *values or confidence intervals*

If we have the data we can either work out the mathematics of the confidence distribution ourselves, or reverse engineer confidence levels for hypotheses from the confidence interval routines in standard packages. However, even in the absence of the data it is still possible to make a reasonable estimate if we have either a confidence interval or a *p* value. If we assume that the confidence distribution is a normal distribution (or a *t* distribution, but this usually makes little difference), it is possible to use the information we have to reverse engineer the confidence distribution and use it to derive whatever confidence levels we require. The assumption of normality might seem suspect, but many distributions are roughly normal, and normal approximations are widely used in statistics even when they are not very close (e.g. the for binomial or Poisson distributions).

As an example, suppose we wanted to estimate the confidence level for the mean of Population A being at least one unit more than B: Table 2, based on the full data set, shows this is 17% based on the small samples. However, if we *only* knew that the sample value of the difference is 0.3, and the *p* value is 67.3%, the online supplement material spreadsheet CLIP.xls is designed to estimate confidence levels. The confidence level in question comes to 16% using the normal distribution—which is close enough to be useful in practice. Alternatively, we could use the 95% confidence interval (−1.2 to 1.8) to derive a similar result from the same spreadsheet.

### *Method 4: bootstrapping*

Bootstrapping is a method of deriving confidence distributions introduced by Efron (1979). It is one of the approaches discussed by Xie and Singh (2013) in their review of confidence distributions; there is a more elementary explanation in Wood (2004). Bootstrapping relies on computer simulation which can estimate approximate probabilities without detailed mathematical models. The online supplement material spreadsheet, resamplepopsab.xlsx, is set up to estimate confidence levels from the data in the small sample in Table 1 by bootstrapping. Because bootstrapping is a simulation method, it will produce a slightly different distribution each time it is run. Figure 2 shows one such bootstrapped confidence distribution: this models the same distribution as the bold line in Figure 1.

When I used the spreadsheet, which analyzes 1000 simulations (resamples) each time, the results for the confidence level of the mean of Population A being at least one unit more than B ranged from 14% to 17%. This is close[2] to the answer of 17% from the confidence distribution in Figure 1.

One of the strengths of bootstrapping is its flexibility in terms of the problems it can model. The spreadsheet used for Figure 2 can easily be adapted to analyze other statistics such as regression and correlation coefficients. Wood (2013) describes how bootstrapping can be used to estimate a confidence level (65%) for a hypothesis of an inverted U shape between two variables. The original paper on whose data this analysis is based (Glebbeek and Bax, 2004) merely produced two nonsignificant *p* values for the two parameters of a quadratic function.

## How should we describe the strength of the evidence based on a sample: confidence levels or *p* values?

Sometimes *p* values are a sensible approach. In an experiment to detect telepathy described in Wood (2016: 4), a volunteer chooses a card at random from a pack of 50 cards, concentrates hard on it, and then another volunteer tries to work out what the chosen card is without being able to see the card or communicate with the first volunteer in any way except, possibly, by telepathy. If the second volunteer gets the right answer, this represents a *p* value of 1/50 or 2% based on the null hypothesis that the second volunteer is guessing. There is no confidence distribution here, and no obvious easier way of coming up with a probability for telepathy than the use of Bayes' theorem, which requires a prior probability for telepathy. The *p* value (2%) is a simple and natural way to summarize the conclusion about the strength of the evidence. But this example is unusual: in many cases, such as the example about Oscar winners described in the "Introduction" above, there is a confidence distribution, and *p* values are neither simple nor natural.

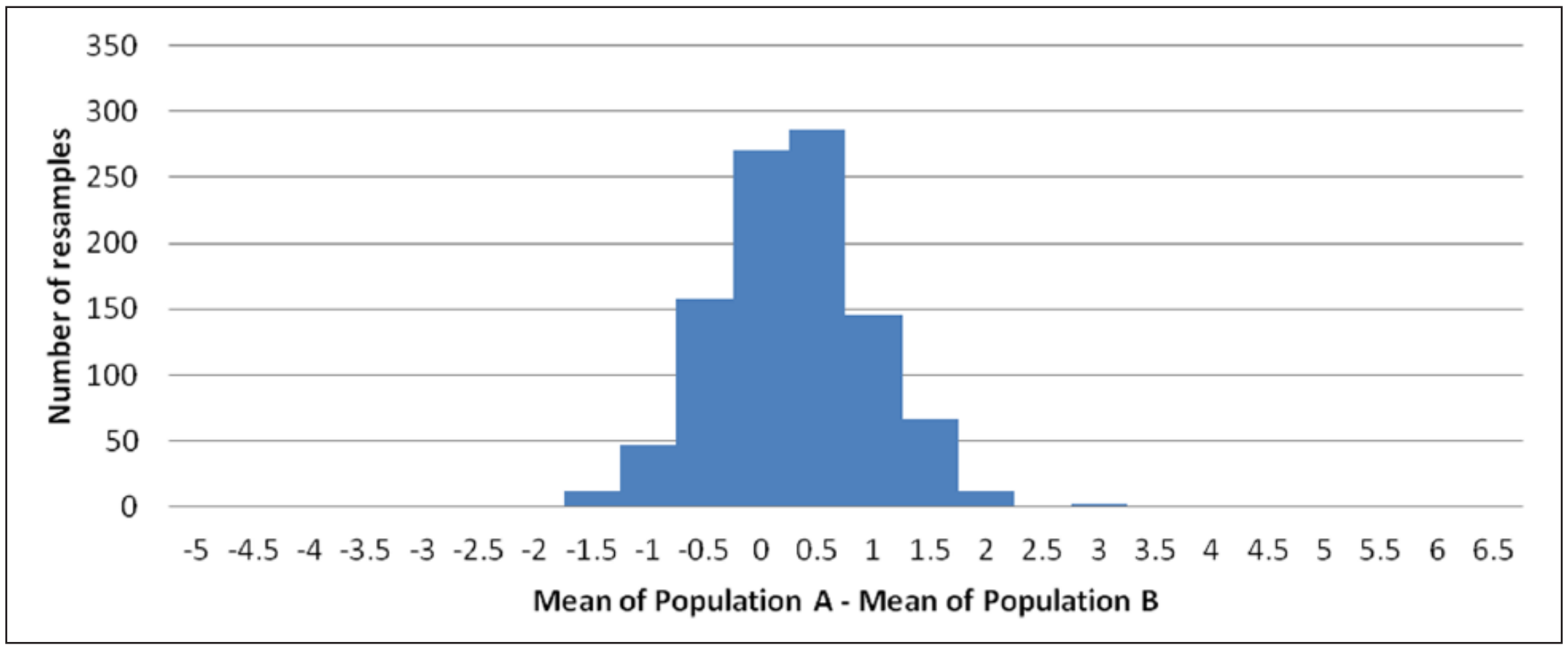

**Figure 2.** Bootstrapped confidence/tentative probability distribution.

Confidence intervals are now widely cited in medical journals: for example, Wallis et al. (2017) concluded that "patients treated by female surgeons were less likely to die within 30 days (adjusted odds ratio 0.88; [95% confidence interval] 0.79 to 0.99, $p=0.04$), but there was no significant difference in readmissions or complications." The confidence interval here gives an indication of the likely error but the *p* value is still the statistic given to quantify the strength of the evidence for the hypothesis about the difference between male and female surgeons. And the phrase "no significant difference" implies that it is the difference which is not "significant," whereas it is actually the *evidence for* the difference which is not significant. Phrases like "significant difference" are likely to reinforce the common misunderstanding that a significant difference is a big or important difference.

In many other disciplines, *p* values remain the sole statistic used to quantify uncertainty due to sampling error; confidence intervals are not given. We have seen in section "Method 4: Bootstrapping" above how Glebbeek and Bax (2004) could only give two nonsignificant *p* values in support of their hypothesis of an inverted U shape between two variables. They also looked at a linear regression model between the same two variables and concluded that "a 1% increase in turnover equals a loss of 1780 Dutch guilders [this was before the introduction of the Euro] … From a management point of view, this is quite substantial" (p. 283). However, the only statistic quoted to quantify the strength of the evidence for this is $p < 0.01$. The confidence interval (−3060 to −500: see Wood, 2013) is not given.

I think there is a very strong case for using confidence levels for the hypotheses of interest to the research in all these papers. So instead of the *p* value quoted above for the comparison between male and female surgeons, we could (using Method 2 above) write:

> Patients treated by female surgeons were slightly less likely to die within 30 days (confidence level: 98%).

And instead of the potentially misleading statement that "there was no significant difference in readmissions or complications," the fact that there was a difference but that the evidence that this would apply to a wider population was weaker could be acknowledged by stating a confidence level (about 90% for the readmission rate).

Or, taking a different perspective, we might feel that, given the likely inaccuracies of the matching and adjustment processes, a difference in the readmission rates of less than 10% is not meaningful. Method 3 can be used to estimate the confidence level for the difference being less than 10%: this comes to about 98% (by estimating the probability of the risk ratio—which is identical to the odds ratio in this case—being $<0.9$ and $>1.1$ and subtracting these two probabilities from 100%). This represents strong evidence that there is not a meaningful difference between male and female surgeons.

The conclusions in Glebbeek and Bax (2004) could be presented in a similar way. The confidence level for increasing turnover leading to a loss is 99.7%. The evidence for the inverted U shape cannot be adequately measured by *p* values: the confidence level of 65% makes the conclusion far clearer than two nonsignificant *p* values.

## Conclusion

Despite being an obvious extension to the idea of confidence intervals, confidence levels such as those above are never cited. As far as I can see, the idea is never even considered, despite the fact that it gives clear answers to questions about the strength of evidence, whereas the ubiquitous *p* values are widely misinterpreted and often fail to provide much useful information. Instead of qualifying the conclusion that patients treated by female surgeons were slightly less likely to die within 30 days with the statement $p=0.04$ (Wallis et al., 2017), we could qualify it with by writing confidence level = 98%.

To summarize, the advantages of citing confidence levels instead of *p* values (when this is possible) are that (unlike *p* values):

- They give a direct, although tentative, estimate of the probability of a hypothesis being valid.
- We can estimate confidence levels for a range of different hypotheses.
- This allows the size of the effect to be taken into account. For example, Table 2 shows that the confidence level for the size of the difference between the two population means being less than one is 79% based on the small sample, and 100.0% based on the large sample.
- The lack of a threshold like 5%, and the term "significant," both of which may encourage people to assume a dichotomy between proven and not proven, may encourage a more nuanced view of statistical conclusions.

But on the negative side, confidence levels, like *p* values, only take account of sampling errors; other sources of error are ignored. And just like confidence intervals, they depend on the assumptions behind the model (*t* distribution, bootstrap model, or whatever) used to derive the confidence distributions. They should be regarded as tentative not definite. And although the *meaning* of assertions like "confidence level = 98%" should be reasonably intuitive, the *method* used to compute it is inevitably more complicated (the bootstrap method is probably the most transparent).

However, confidence intervals are used, despite their tentative status, and they are inevitably likely to be the basis of informal calculations of confidence levels such as those I am suggesting here ("The confidence interval does not include zero so …"). Why not make these informal calculations explicit?

The answer may be that doing so would make the conclusions clearer, which would expose them to more scrutiny. If, for example, only 50% of hypotheses which achieve a confidence level of 95% or more turn out to be true, this would suggest something is wrong with the assessment of confidence levels.

I have also argued that confidence levels could be regarded as tentative probabilities: tentative because their derivation inevitably depends on assumptions about which we cannot be certain (e.g. a flat prior distribution). Confidence seems an unnecessary concept in an already overcrowded conceptual framework. It is also a bad name for a concept in which we should not have too much confidence. *Tentative probability* seems a better term because it connects with intuitions about probability and acknowledges its limitations. However, in the short term, keeping the word confidence may be helpful because it gives the idea of tentative probabilities for hypotheses the support of an established, and so relatively unquestioned, theoretical framework.

### Declaration of conflicting interests

The author(s) declared no potential conflicts of interest with respect to the research, authorship, and/or publication of this article.

### Funding

The author(s) received no financial support for the research, authorship, and/or publication of this article.

### Supplemental material

Supplemental material for this article is available online.

### Notes

1. This is a two-tailed *p* value. As the researcher's hypothesis is directional (the mean of A will be more than B), there might be a case for using a one-tailed test which would give a *p* value of 0.33. However, most research cites only two-tailed tests, and SPSS gives no option for one-tailed tests for the difference of two means, so I will stick with the two-tailed test here. One-tailed tests give lower *p* values, indicating *more* significant results: they are a reasonable approach where there is a clear directional hypothesis *before* the date is collected. However, their use may lead to the suspicion that the researcher has formulated a directional hypothesis *after* seeing the data as a way of obtaining a more significant result. Two-tailed tests avoid this problem, so significant results are more secure. In practice, in this example, it makes little difference: the result for the small samples is not significant in either case, and for the large samples (see Table 1) the result is significant with one ($p=0.002$) or two tails ($p=0.004$).
2. But not identical: the bootstrap estimate is slightly less, possibly because small samples are less likely to include extremes and so may underestimate the variability in the population. However, worrying unduly about the precision of the confidence levels seems unreasonable because the sample is small, so everything is just a rough guess.

### ORCID iD

Michael Wood https://orcid.org/0000-0003-2119-0942

## Author biography

Michael Wood is a Visiting Fellow at Portsmouth University.